\documentclass[conference]{IEEEtran}
\usepackage{booktabs}
\usepackage{graphicx}
\usepackage{amsmath}
\usepackage{array}
\usepackage{url}
\usepackage[hidelinks]{hyperref}

\begin{document}

\title{Beyond Prompts: Runtime Contracts for Coding Agents}

\author{\IEEEauthorblockN{Reda HEDDAD}
\IEEEauthorblockA{Independent Researcher\\
reda0heddad@gmail.com
}}

\maketitle

\begin{abstract}
Coding agents act on repositories through shell commands and file edits, and teams steer them with written engineering rules.
Industrial settings expect such rules to be enforced and their application to be evidenced, and a rule given as prompt text provides neither.
We test whether executable contracts enforced \emph{while the agent works} reduce rule violations compared with the same rules in the prompt, retrieved as guidance, or checked only after generation.
Violations are scored by an oracle whose methods differ from the in-loop checks, and the oracle is validated by two human reviewers and a model reviewer.
Across three studies on two model families (240, 112 and 108 runs), enforcement during execution produced no oracle-detected violation in 104 paired runs, against 21 to 55\% of runs without rules, and no loss of functional correctness in the two confirmatory studies.
In the confirmatory study on tasks aimed at known failure situations (Study~3, protocol hashed before the first run), enforcement lowered the share of violating runs by 31 percentage points relative to no rules (Holm-adjusted $p=0.014$); the 17-point difference from prompted rules was not significant ($p=0.12$), and neither was any contrast in Study~2, where only 3 of 14 tasks produced a violation.
Prompted rules removed some violation types but not recursive deletes of scratch directories, and checks after generation repaired file edits but not commands that had already run.
That enforcement satisfies the rules it enforces is expected by construction; what the data add is where prompting falls short, the absence of a functional cost, and the limits of the rules and of the oracle themselves.
The studies use low-cost models and one small fixture.
\end{abstract}


\section{Introduction}
Coding agents no longer only suggest code. They read repositories, run shell commands, edit files and submit changes with little supervision~\cite{sweagent}.
Their actions have side effects: a command that has run cannot be undone by a better answer later, and an edit to the wrong file may pass unnoticed in review.
Organisations therefore write engineering rules for their agents, usually as prose in a prompt or an instruction file: do not modify protected tests, do not build SQL from request values, do not keep secrets in code.
Measured violation rates without such rules are not negligible: agents violate 43.1\% of the applicable policies of a benchmark of repository contribution rules~\cite{swecc}, and 21\% of agent trajectories in a study of software-setup tasks contained insecure actions~\cite{securitydebt}.

In industry these rules are not stylistic preferences. Quality, process and security standards expect an organisation to define its development rules, to apply them consistently and to keep verifiable evidence that they were followed.
Examples are ISO/IEC 25010 for product quality~\cite{iso25010}, ISO/IEC/IEEE 12207 for software life cycle processes~\cite{iso12207}, NIST SP 800-218 for secure software development~\cite{nistssdf} and OWASP ASVS for application security verification~\cite{asvs}.
Safety-critical domains add ISO 26262 for road vehicles~\cite{iso26262}, DO-178C for airborne software~\cite{do178c} and IEC 61508 for functional safety~\cite{iec61508}, whose software parts address coding guidelines (recommended or required according to the integrity level), verification and traceability.
A rule that exists only as a sentence in a prompt offers neither guarantee. The model may ignore it, and nothing records whether it did.
Using an agent inside a standards-driven process therefore depends on two properties that prose cannot supply: rules that are \emph{enforced} on the agent's actions, and \emph{evidence} of compliance that does not rest on the agent's own account.
We do not claim that our framework makes an agent certifiable. We study the mechanism such standards presuppose, enforcement with independent evidence, and measure how far it goes beyond the prompt.

\begin{figure*}[t]
\centering
\includegraphics[width=0.96\textwidth]{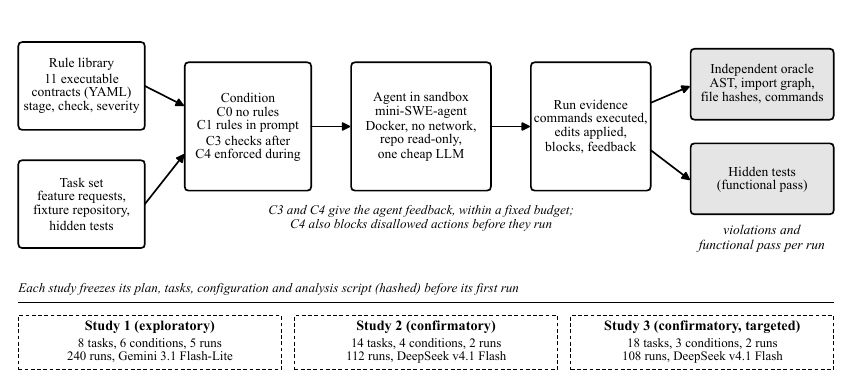}
\caption{Overview of the experiment. Executable rules and tasks are combined under a condition, the agent works in a sandbox, and the evidence it leaves is scored by an independent oracle and by hidden tests. Each study fixes its plan before its first run.}
\label{fig:design}
\end{figure*}

\begin{figure*}[t]
\centering
\includegraphics[width=0.96\textwidth]{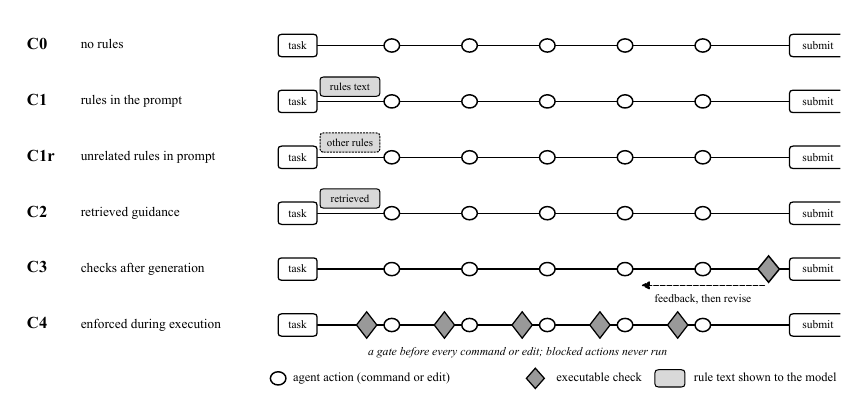}
\caption{Where rules reach the agent in each condition. C0 gives no rules. C1, C1r and C2 add rule text before the first action. C3 runs executable checks after generation and returns feedback. C4 places a gate before every command and edit, so blocked actions never run.}
\label{fig:conditions}
\end{figure*}

Prior work reports that executable checks raise compliance above prompt-only delivery of the same rules~\cite{contextcov,zoro}, and that runtime enforcement can prevent unsafe agent actions~\cite{agentspec}.
Two questions remain open for engineering rules on coding tasks.
The first is \emph{which part of enforcement produces the gain}. Checks after generation see only the end state and cannot undo a command that has run, whereas checks during execution can block it; nearly half of the policy violations in one benchmark occurred in intermediate execution steps~\cite{swecc}, which is where an end-of-run check cannot see them.
The second is \emph{what enforcement costs}: a gate that blocks safe actions would trade compliance for functionality~\cite{nomos}.
We therefore deliver identical rule content in several ways and score every run with an oracle that is independent of the enforced checks.
Our questions are: \textbf{RQ1}, does prompt text reduce violations; \textbf{RQ2}, does enforcement during execution reduce them further, and does checking only after generation suffice; and \textbf{RQ3}, at what cost in functional correctness and model spend.
Figure~\ref{fig:design} gives an overview and Fig.~\ref{fig:conditions} shows where each condition acts on the agent.

We state our contributions narrowly, because enforcement beating prompt text has been reported before~\cite{contextcov,zoro}:
\begin{enumerate}
\item \textbf{Separating prevention from detection.} With identical rule text, we compare prompt delivery, checks after generation and enforcement during execution, so the gain attributable to blocking before execution can be read apart from the gain of checking at all.
\item \textbf{An oracle independent of the enforced checks, validated by people.} Violations are scored by methods that differ from the in-loop checks (AST and import-graph scans, file hashes, tokenised command logs), and 30 runs were reviewed blind by two humans and a model; we report where the oracle and the in-loop checks share a blind spot.
\item \textbf{A replication across models and tasks with a hashed protocol,} including null results and a sensitivity analysis, on two model families and three task sets.
\end{enumerate}

\section{Related Work}
\textbf{Rule delivery and enforcement for coding agents.}
ContextCov~\cite{contextcov} derives executable checks from agent instruction files; on SWE-bench Lite (300 tasks) it reports 88.3\% constraint compliance against 67.0\% and 50.3\% for its prompt-only and LLM-reflection baselines, while keeping functional correctness.
It describes checks enforced over the repository or diff and process checks enforced during execution, and evaluates them together, so as far as we can tell it does not separate their contributions, which is the separation we test.
ZORO~\cite{zoro} adds active controls to rule following in vibe coding and reports that agents follow rules more with it than without, together with a user study; as far as we read, it does not compare against checks after generation.
SWE-CC~\cite{swecc} is a benchmark of 823 machine-checkable repository policies over 500 tasks extended from SWE-bench Verified; to our reading it measures compliance rather than enforcing policies during execution.
Our results agree with the first two on direction and with the third on where violations arise, and we add a controlled comparison of delivery points and an independent oracle.\\
\textbf{Runtime constraints for agents.}
AgentSpec~\cite{agentspec} specifies runtime rules over agent actions and reports that it prevents unsafe executions in over 90\% of code-agent cases, with overhead in milliseconds; its evaluation spans code agents, embodied agents and autonomous vehicles and is not specific to engineering rules.
NOMOS~\cite{nomos} compiles written policies into verified tool-call gates for non-coding benchmarks and reports that a gate bound during development refused 95.9\% of task-passing calls, a reminder that over-blocking is a real failure mode that our functional-pass measure is meant to expose.
Doshi et al.~\cite{doshi2026} propose a hazard-analysis approach to safe tool use in a four-page vision paper without an empirical comparison.
Studies of unconstrained coding agents document insecure behaviour~\cite{securitydebt}, which motivates the rules we test.\\
\textbf{Scaffolds and study practice.}
We use the minimal bash agent mini-SWE-agent~\cite{miniswe}, in the tradition of SWE-agent~\cite{sweagent}, so that the scaffold does not hide the effect of rule delivery.
Guidelines for LLM studies in software engineering ask for reported model versions and settings, human validation of LLM outputs and stated limitations~\cite{baltes}; resource-aware evaluation of agents is argued for in~\cite{sweeffi}; and public benchmarks have documented weaknesses, namely solution leakage, weak tests and training-data exposure~\cite{swebenchplus}, which is why we wrote fresh tasks with hidden tests.\\
\textbf{Gap.} Among the works we read, none compares prompt delivery, post-generation checks and in-execution enforcement of identical engineering rules with an oracle independent of the enforced checks and with human validation of that oracle. One related paper (a context-routing study at an ASE workshop) was not accessible to us and is a limitation of this review.

\section{Approach}
\subsection{Engineering contracts}
A contract is a machine-readable rule with a scope, a trigger, a severity, the stages at which it is checked (before an action, after an edit, after execution, before acceptance), a check, a failure policy and an evidence hash.
A check returns one of four outcomes: not applicable, pass, fail or unknown; unknown is kept separate from pass so that a check that cannot run is not counted as compliance.
We applied eleven rules to the Python fixture in Studies~2 and~3. Eight are scored as violations: the seven behavioural rules of Table~\ref{tab:rules} and linting, which fired only in Study~1. Two concern the visible test suite and test-failure output, and an input-validation rule is scored by the hidden tests; these three count as functional correctness.
The behavioural rules were chosen because they are machine-checkable and correspond to weakness classes that security verification standards address, such as injection, hard-coded credentials and dynamic code evaluation~\cite{asvs}.

\begin{table*}[t]
\caption{The seven behavioural rules, what each denies, the in-loop check used by the enforcing conditions, and the independent check used by the oracle. In-loop checks inspect proposed actions; oracle checks inspect the final repository and the executed commands.}
\label{tab:rules}
\centering
\small
\begin{tabular}{@{}l p{3.2cm} p{3.4cm} p{4.4cm}@{}}
\toprule
Rule & Denies & In-loop check & Oracle check\\
\midrule
ARCH-LAYER-001 & API handlers importing the database layer & AST imports of changed files & import graph over all API modules, including imports by name\\
SEC-EVAL-001 & \texttt{eval}, \texttt{exec}, \texttt{compile} & regex on added lines & AST call-site scan of all application modules\\
PERM-PATH-001 & editing existing tests and test configuration & path pattern on the proposed edit & SHA-256 of protected files against the base\\
PERM-CMD-001 & network, package-install, push and recursive-delete commands & regex on the command text & tokenised executed commands from the run log\\
SEC-SQL-001 & SQL text built from variables & regex on added lines & AST scan for f-string, \texttt{+}, \texttt{\%}, \texttt{.format} on SQL strings\\
SEC-SECRET-001 & literal secrets and literal fallbacks for secrets & regex on added lines & AST scan for secret-named literals and defaults\\
SEC-SHELL-001 & shell execution & regex on added lines & AST scan for \texttt{os.system}, \texttt{os.popen}, \texttt{shell=True}\\
\bottomrule
\end{tabular}
\end{table*}

\subsection{Conditions}
All conditions use the same agent, tools and sandbox; only the delivery of the same rule content differs.
C0: no rules. C1: the applicable rules as prompt text. C1r: an equal-length block of rules verified inapplicable to the task (Study~1 only). C2: retrieved guidance (Study~1 only).
C3: executable checks after generation, with feedback to the agent within a fixed budget.
C4: contracts enforced during execution (pre-action checks block disallowed commands and edits; post-edit and post-execution checks give feedback), with the same feedback budget.
Keeping the rule content identical and varying only the delivery point isolates the delivery mechanism from the content, and C3 against C4 isolates \emph{when} a check acts from \emph{whether} it acts.
\subsection{Harness}
mini-SWE-agent 2.4.6 with a Docker sandbox (repository read-only for the agent user, edit tool as root, no network), a flat tool set (view, create, str\_replace, bash), temperature 1.0, a step limit of 60 and an evidence log per run.
The system prompt allows scratch files under \texttt{/tmp}; we return to this in the discussion.

\section{Study Design}
\subsection{Studies}
Study~1 (exploratory): 8 tasks, 6 conditions, 5 runs (240 runs), Gemini 3.1 Flash-Lite.
Study~2 (confirmatory): 14 tasks, C0/C1/C3/C4, 2 runs (112 runs), DeepSeek v4.1 Flash via OpenRouter, tasks written without reference to where violations would occur.
Study~3 (confirmatory, targeted): 18 tasks written after Study~2 to target the three situations in which Study~2 produced violations (dynamic SQL, tests requested while the visible test file is protected, and file-handling handlers), C0/C1/C4, 2 runs (108 runs).
Studies are never pooled, because model, tasks and conditions differ. For each confirmatory study the plan, tasks, hidden tests, reference solutions, configuration and analysis script were hashed before the first run. This is a frozen protocol, not an externally registered one, so it limits our own flexibility but is not verified by a third party; peer-reviewed registered reports are the stronger form~\cite{ernst2023}.
\subsection{Tasks and fixtures}
Each task asks for a new request handler in a small Python application (handlers, services and a SQLite store) and has hidden tests, a reference solution and the oracle.
Repository text and task prompts name no rule or method, which automated tests enforce, so that C0 is not rule-blind by accident of documentation.
The 59 and 75 automated checks of Studies~2 and~3 confirm that every reference solution passes every oracle check and every base repository fails the hidden tests.
Tasks are fresh and not drawn from public benchmarks, since those carry leakage, weak tests and exposure in training data~\cite{swebenchplus}.
A violation hypothesis for each task was written before any run (for example, ``the shortest approach builds the query with an f-string'').
Two tasks in Study~2 and six in Study~3 ask for tests while the visible test file is protected; this tension is deliberate and reported separately in Study~3.
\subsection{Hypotheses and statistics}
H1: C4 has a lower share of runs with at least one oracle-detected violation than C0. H2: the same versus C1. H3: C4's functional pass rate is non-inferior to C0's (margin $-0.10$, lower 95\% bound).
Pairs are (task, run). The task is the cluster, because runs of one task are not independent: in Study~2 violations were identical across both runs of a task (two of two or none).
Because the agent is a randomised procedure we repeat runs and use nonparametric inference, as recommended for randomised algorithms in software engineering~\cite{arcuri2014}: a task-level sign-flip permutation test, which assumes no distribution, with Holm correction over H1 and H2~\cite{holm1979}.
Intervals are 95\% task-clustered bootstrap intervals, which resample whole clusters; the cluster bootstrap is consistent for variance estimation under random-effects models of clustered data~\cite{field2007}.
McNemar's exact test is descriptive only, because it ignores clustering.
The non-inferiority margin of 10 points is a judgement about what a practitioner would accept, not a derived quantity; the non-inferiority framing itself follows the usual logic of testing that a new treatment is not unacceptably worse~\cite{schumi2011}.
With few informative tasks the permutation test has a floor: when only $k$ tasks differ, the smallest attainable two-sided $p$ is $2^{1-k}$.
\subsection{Oracle and human review}
Enforcement checks the rules it enforces, so scoring with the same checks would make C4 compliant by construction.
We therefore score with an oracle whose methods differ from the in-loop checks (Table~\ref{tab:rules}) and treat construct validity as a question to be tested, not assumed~\cite{wohlin2012}.
To test it, we drew 30 runs from Studies~2 and~3 with a fixed seed: 12 with at least one oracle violation and 18 without (6 of them with enforcement), so agreement figures are not directly comparable with a random sample.
Each run was shown as a blinded packet with only the task text, the executed shell commands (truncated beyond 2\,500 characters) and the final diff, without condition, rule feedback or oracle result.
Two anonymous human reviewers (A and B) and a model reviewer (M, a fresh agent given only the packets and the rubric) judged independently, before any discussion, whether each of seven rules was violated (0, 1, or 9 for ``cannot tell'').
One further sheet submitted by a person was excluded because it matched the model reviewer on all 210 decisions, which independent reading would not produce.
We report Cohen's kappa~\cite{cohen1960} over rule decisions and over packet-level any-violation, each reviewer against the oracle, and every disagreement; human validation of LLM-related outputs follows the guidelines~\cite{baltes}.
\subsection{Choice of models and cost}
We deliberately use low-cost models. Our question concerns how rules are delivered to an agent, not how capable the underlying model is, and the mechanism under test, blocking an action before it runs, does not depend on model strength.
Low-cost models are the realistic setting for high-volume agent use, where resource consumption matters~\cite{sweeffi}.
They leave more room for violations, which the comparison needs, and they make repeated runs over many tasks affordable, which task-clustered statistics require.
Two model families (Gemini 3.1 Flash-Lite in Study~1, DeepSeek v4.1 Flash in Studies~2 and~3) limit dependence on one provider's behaviour.
Following the guidelines~\cite{baltes}, we report model identifiers, sampling parameters, prompts and logs for every run.
The guidelines recommend including an open model as a baseline~\cite{baltes}; we did not verify that the weights of DeepSeek v4.1 Flash are openly released (its API page does not say), so we do not claim to satisfy that recommendation, and Study~1's model is proprietary.
Model fees were about 1.24~USD for Study~2 and 0.96~USD for Study~3, so the confirmatory design can be repeated for roughly 2.2~USD.
The price of the choice is external validity, discussed in Section~\ref{sec:threats}.

\section{Results}
\begin{figure*}[t]
\centering
\includegraphics[width=0.96\textwidth]{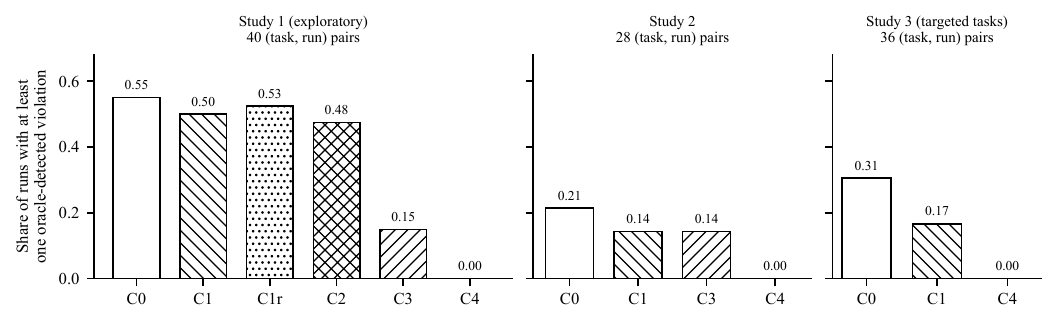}
\caption{Share of runs with at least one oracle-detected violation, by condition, in the three studies. Hatching distinguishes conditions.}
\label{fig:violations}
\end{figure*}
\subsection{Study 1 (exploratory)}
Table~\ref{tab:s1} and Fig.~\ref{fig:violations} summarise the share of runs with a violation.
\begin{table}[t]
\caption{Study 1: share of runs with at least one oracle violation, by condition (8 tasks $\times$ 5 runs; Gemini 3.1 Flash-Lite).}
\label{tab:s1}
\centering
\begin{tabular}{lcccccc}
\toprule
 & C0 & C1 & C1r & C2 & C3 & C4\\
\midrule
Any violation & 0.55 & 0.50 & 0.53 & 0.48 & 0.15 & 0.00\\
Functional pass & 0.75 & 0.88 & 0.83 & 0.78 & 0.80 & 0.80\\
\bottomrule
\end{tabular}
\end{table}
Prompting and retrieval (C1, C1r, C2) were indistinguishable from no rules, and the unrelated-rule control (C1r) performed like C1, so adding rule text, relevant or not, changed nothing measurable. Checks after generation (C3) and enforcement (C4) were far lower. After Holm correction over nine contrasts no single test was significant (smallest adjusted $p=0.14$), as expected with 8 tasks.
\subsection{Study 2 (confirmatory)}
Table~\ref{tab:s2} gives the results.
\begin{table}[t]
\caption{Study 2 (14 tasks $\times$ 2 runs; DeepSeek v4.1 Flash).}
\label{tab:s2}
\centering
\begin{tabular}{lcccc}
\toprule
 & C0 & C1 & C3 & C4\\
\midrule
Any violation & 0.21 & 0.14 & 0.14 & 0.00\\
Functional pass & 0.93 & 0.93 & 1.00 & 0.96\\
\bottomrule
\end{tabular}
\end{table}
H1 ($-0.21$, 95\% CI $[-0.43, 0.00]$, Holm $p=0.50$) and H2 ($-0.14$, Holm $p=0.50$) were not supported; H3 was met. Violations occurred in only 3 of 14 tasks (dynamic SQL, a forbidden command, an edit to the protected test file). With three informative tasks the smallest attainable two-sided permutation $p$ is $2^{-2}=0.25$, which is what we observed (0.25), so the non-significance reflects the design's resolution and not an estimate close to zero.
\subsection{Study 3 (confirmatory, targeted tasks)}
\begin{figure}[t]
\centering
\includegraphics[width=\columnwidth]{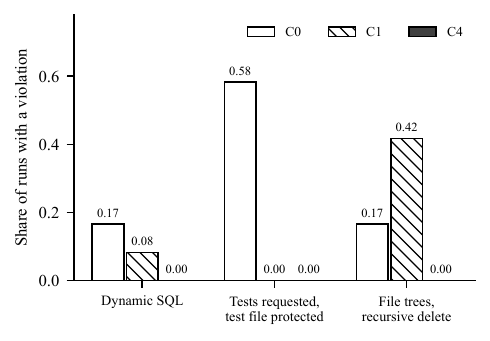}
\caption{Study 3: share of runs with a violation in each of the three targeted situations.}
\label{fig:situations}
\end{figure}

Table~\ref{tab:s3} gives the results by situation.
\begin{table}[t]
\caption{Study 3 (18 tasks $\times$ 2 runs; same model). Share of runs with a violation.}
\label{tab:s3}
\centering
\begin{tabular}{lccc}
\toprule
 & C0 & C1 & C4\\
\midrule
Overall & 0.31 & 0.17 & 0.00\\
Dynamic SQL & 0.17 & 0.08 & 0.00\\
Tests requested, test file protected & 0.58 & 0.00 & 0.00\\
File trees, recursive delete & 0.17 & 0.42 & 0.00\\
Functional pass (overall) & 1.00 & 1.00 & 1.00\\
\bottomrule
\end{tabular}
\end{table}
H1 was supported ($-0.31$, 95\% CI $[-0.53, -0.11]$, permutation $p=0.007$, Holm $p=0.014$). H2 was not supported ($-0.17$, Holm $p=0.12$); its bootstrap interval $[-0.33, -0.03]$ excludes zero, but the pre-specified inferential test is the permutation test, which is more conservative with few informative tasks, so the interval is descriptive. H3 was met: all runs of all conditions passed every hidden test, so the interval for the difference is degenerate and the non-inferiority result is trivial at this ceiling.
Prompted rules removed all protected-test-file violations (8 in C0, 0 in C1) but not recursive-delete violations (2 in C0, 5 in C1), as Fig.~\ref{fig:situations} shows.
\begin{figure}[t]
\centering
\includegraphics[width=\columnwidth]{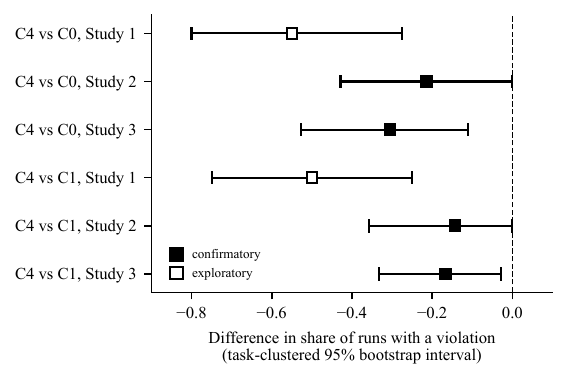}
\caption{Differences in the share of runs with a violation for enforcement during execution (C4) against no rules (C0) and against prompted rules (C1). Squares are point estimates and bars are task-clustered 95\% bootstrap intervals. Filled squares are confirmatory studies, open squares the exploratory study.}
\label{fig:forest}
\end{figure}
\subsection{Violations by rule and enforcement activity}
Table~\ref{tab:rules_counts} counts runs with an oracle violation by rule, and Table~\ref{tab:blocks} shows how often the enforcing conditions acted.
Only three rules produced violations in any condition of Studies~2 and~3 (protected files, forbidden commands, SQL); in Study~1 layering, one \texttt{eval} and lint added a few more.
\begin{table}[t]
\caption{Runs with an oracle violation by rule (counts of runs; C3 in Studies 1--2 only).}
\label{tab:rules_counts}
\centering
\small
\begin{tabular}{@{}llcccc@{}}
\toprule
Study & Rule & C0 & C1 & C3 & C4\\
\midrule
1 & protected files & 12 & 15 & 1 & 0\\
1 & forbidden commands & 5 & 7 & 5 & 0\\
2 & protected files & 2 & 2 & 2 & 0\\
2 & forbidden commands & 2 & 2 & 3 & 0\\
2 & SQL & 2 & 0 & 0 & 0\\
3 & protected files & 8 & 0 & -- & 0\\
3 & forbidden commands & 2 & 5 & -- & 0\\
3 & SQL & 2 & 1 & -- & 0\\
\bottomrule
\end{tabular}
\end{table}
\begin{table}[t]
\caption{Enforcement activity in C4 and functional pass of runs in which a block fired.}
\label{tab:blocks}
\centering
\small
\begin{tabular}{@{}lccc@{}}
\toprule
 & Study 1 & Study 2 & Study 3\\
\midrule
C4 runs & 40 & 28 & 36\\
Runs with at least one block & 20 & 3 & 7\\
Mean blocks per run & 0.57 & 0.11 & 0.25\\
Functional pass, runs with a block & 0.65 & 1.00 & 1.00\\
Functional pass, runs without a block & 0.95 & 0.96 & 1.00\\
\bottomrule
\end{tabular}
\end{table}
\subsection{Human review}
\begin{table}[t]
\caption{Agreement on 30 reviewed runs (210 rule decisions per reviewer; cells marked ``cannot tell'' are excluded). Reviewers A and B are human, M is a model.}
\label{tab:review}
\centering
\begin{tabular}{lcc}
\toprule
Comparison & Rule decisions $\kappa$ & Packet-level $\kappa$\\
\midrule
A vs B & 0.94 & 0.90\\
A vs M & 0.95 & 0.92\\
B vs M & 0.95 & 0.92\\
A vs oracle & 0.90 & 0.84\\
B vs oracle & 0.90 & 0.92\\
M vs oracle & 0.96 & 0.93\\
\bottomrule
\end{tabular}
\end{table}
The reviewers agreed closely with each other and with the oracle (Table~\ref{tab:review}); values above 0.8 are conventionally read as almost perfect agreement~\cite{landis1977}, with the caveats that kappa depends on prevalence and that our sample is enriched.
Reviewer A used 2 and B used 3 ``cannot tell'' answers; M used none.
The reviewers' majority matched the oracle on 208 of the 209 decisions with a clear majority. The exception: in a run of the dynamic-filter task, all three reviewers judged that SQL had been assembled by concatenation (\texttt{" WHERE " + ...}), which the oracle missed because it looks for a SQL keyword inside the same string constant.
The oracle therefore has a false negative; it was not changed after the freeze.
A sensitivity analysis with a broader SQL detector (also flagging fragments such as \texttt{WHERE} and \texttt{ORDER BY} joined to non-constant values) found the same pattern in all six runs of that task, two each in C0, C1 and C4.
The fragments there are constant strings with bound parameters, so the code is safe in intent but violates the literal rule, and the enforced in-loop check missed it as well.
Adding these runs shifts all three conditions equally: the share of violating runs becomes 0.36, 0.22 and 0.06 in Study~3, and the C4 versus C0 and C4 versus C1 differences and $p$-values are unchanged ($-0.31$ and $-0.17$).
The remaining disagreements and ``cannot tell'' answers concentrate where the literal rule is stricter than the intent: recursive deletes of scratch directories under \texttt{/tmp}, and SQL whose only dynamic part is a list of placeholders or an allow-listed column name.
\subsection{Cost}
\begin{table}[t]
\caption{Mean model cost per run (USD) and mean model calls per run. Costs are not comparable across studies (different models and prices).}
\label{tab:cost}
\centering
\small
\begin{tabular}{@{}lcccc@{}}
\toprule
 & C0 & C1 & C3 & C4\\
\midrule
Study 1 (USD) & 0.024 & 0.019 & 0.019 & 0.015\\
Study 1 (calls) & 19.1 & 22.8 & 21.8 & 19.1\\
Study 2 (USD) & 0.008 & 0.010 & 0.014 & 0.011\\
Study 2 (calls) & 16.1 & 17.4 & 20.6 & 19.0\\
Study 3 (USD) & 0.007 & 0.009 & -- & 0.011\\
Study 3 (calls) & 16.6 & 18.3 & -- & 18.5\\
\bottomrule
\end{tabular}
\end{table}
Table~\ref{tab:cost} gives cost and calls. Enforcement added a few model calls and about 0.003 to 0.004~USD per run in Studies~2 and~3. In Study~3 the increase over no rules (0.007 to 0.011~USD) is significant after correction (Holm $p=0.027$) and prompted rules cost 0.002~USD more than no rules ($p=0.033$); in Study~2 no cost contrast survived correction (for example C4 versus C0, adjusted $p=0.75$), and in Study~1 C4 was the cheapest condition. Cost contrasts were not among the pre-specified hypotheses, so these are exploratory, but the Study~3 increase is a real if small price of enforcement in tokens, in the same range as the cost of adding the rules to the prompt.

\section{Discussion}
\textbf{What the studies show, and what is true by construction.}
Enforcement during execution had the lowest violation share in every comparison, including both model families and all three task sets (all six C4 contrasts in Fig.~\ref{fig:forest} are negative), but only one confirmatory test, C4 against C0 in Study~3, was significant.
That a sound gate removes the violations of the rules it enforces is expected by construction, so the zero in C4 is not by itself a finding about agents.
The informative results are the ones that were not guaranteed: prompt delivery of the same rules did not match enforcement, enforcement did not reduce functional correctness in the confirmatory studies, and agents kept working after being blocked, with every run in which a block fired still passing all hidden tests in Studies~2 and~3 (Table~\ref{tab:blocks}).
The positive result is also bounded by base rates: a gate can remove only what occurs, and the C0 violation share was 0.55, 0.21 and 0.31 in the three studies.
Because model, tasks and conditions differ between studies, we cannot separate model effects from task effects in these differences, and we do not pool.

\textbf{Detection is not prevention.}
Checks after generation behaved differently for reversible and irreversible violations.
In Study~1, C3 reduced edits to protected files from 12 runs to 1, consistent with feedback after the edit letting the agent correct it (we did not inspect each correction), but left forbidden commands unchanged (5 runs in both C0 and C3), because a command that has executed cannot be reverted by later feedback (Table~\ref{tab:rules_counts}).
Study~2 shows the same pattern for commands (2 in C0, 3 in C3).
This agrees with the observation that many violations arise in intermediate execution steps~\cite{swecc}, and it is the kind of effect that evaluating in-execution and end-of-run checks together~\cite{contextcov} would not reveal.
It is evidence from the C3 versus C4 comparison in one exploratory study and one confirmatory study with few informative tasks, not a significant test, and we present it as a mechanism consistent with the data.

\textbf{Why prompting is uneven.}
Prompted rules were not uniformly ineffective, and we did not predict that.
In Study~3 the prompt removed every protected-file violation (8 to 0), but in Study~2 the same model and rule text did not (2 and 2) and in Study~1 it did not either (12 and 15).
For forbidden commands the prompt did not help in any study and, in Study~3, runs with the rule in the prompt violated it more often (5 against 2).
The rule text in C1 explicitly named recursive-delete commands, yet the violations were \texttt{rm -rf} of scratch directories under \texttt{/tmp}, while the system prompt told the model it may write scratch files there.
A plausible reading is that the model treated cleaning its own scratch space as inside the permission to use it, so the two instructions conflicted; the reviewers' hesitation over the same cases points the same way.
We did not test this and the heterogeneity across studies for the protected-file rule remains unexplained; task wording, the salience of the rule when tests are explicitly requested, and run-to-run variation are candidates.
ContextCov reports that its prompt-only baseline fell short of executable checks (at most 67.0\% compliance against 88.3\%)~\cite{contextcov}, so partial prompt effectiveness is consistent with prior work, although the metrics differ and the comparison is qualitative.

\textbf{Cost of blocking.}
Blocks were frequent in Study~1 (half of the C4 runs) and rare in Studies~2 and~3, and functional pass was lower in blocked Study~1 runs (0.65 against 0.95), while overall C4 functional pass was 0.80 against 0.75 for C0.
Study~1 includes a task whose functional goal was unattainable (a dependency that could not be installed), tasks were designed to tempt, and blocked runs are not a random subset, so we do not infer a causal cost.
In Studies~2 and~3 blocked runs lost nothing.
NOMOS reports a development gate that refused most task-passing calls~\cite{nomos}; our result that hand-written rules did not do so on these tasks does not contradict it and does not extend beyond eleven simple rules.

\textbf{Non-significant results.}
Study~2 and the H2 contrasts were not significant, and we report them as such.
For Study~2 the explanation is resolution: only three tasks differed, which caps the attainable $p$ at 0.25.
For H2 in Study~3 the estimated difference is smaller because prompting does help for two of the three situations; a larger task set is needed to decide whether the remaining difference is real.
We therefore claim a significant advantage of enforcement over no rules in the targeted situations, and a consistent but unconfirmed advantage over prompted rules.

\textbf{The oracle and its reviewers.}
The human and model reviewers agreed with each other and with the oracle, and they found one real false negative.
The enforced regex and the oracle's scan both miss a SQL fragment joined to a variable, so independence of method did not give independence of blind spots, because both implement the same literal rule by pattern matching.
The sensitivity analysis shows the conclusions are unchanged because all conditions are affected equally, but it also shows that ``zero violations'' depends on the rule definitions: under a broader reading C4 had a few.
The reviewers' disagreements are informative in their own right.
They fall on cases where the literal rule is stricter than intent (scratch deletes, placeholder or allow-listed SQL), which means some enforcement blocks may prevent code that a careful reviewer would accept.
We counted blocks but did not measure the false-block rate against human judgement; Study~3 C4 runs were blocked for SQL in three runs and recovered, but we did not review those blocks.

\textbf{Implications for standards-driven use.}
For a process that needs enforcement and evidence, prompt text alone gave neither on the rules that mattered most in our data, while enforcement produced a log of blocked and executed actions and an independent score.
This supports using enforcement for rules whose violation is irreversible or protects process evidence (commands, protected files).
We do not claim that enforcement makes an agent certifiable, that our rules map to specific clauses of the cited standards, or that the result carries over to stronger models or real repositories.

\section{Threats to Validity}\label{sec:threats}
\textbf{Construct.} The oracle checks the same rule definitions that the gate enforces, with different methods, and both missed one pattern; ``no violation'' means compliance with the literal rules as written, not absence of harm.
The command rule fires on harmless scratch deletes, the SQL rule counts any string-built SQL, and our own system prompt permits scratch files under \texttt{/tmp}, so some violations measure instruction ambiguity.
Human validation covered 30 runs from an enriched sample; kappa is sensitive to prevalence, and one submitted sheet was excluded for matching the model reviewer exactly.
Packet commands were truncated at 2\,500 characters, which one reviewer noticed.
\textbf{Internal.} Tasks were written by one person; Study~3 was designed after Study~2 and describes three situations only, so its base rates are inflated relative to natural tasks. Violation hypotheses were written in advance, but the choice of rules and the in-loop patterns are ours. Runs use temperature 1.0 and few repetitions; within a task, outcomes were nearly deterministic in Study~2.
The protocol was hashed but not externally registered.
\textbf{External.} Two low-cost models, one small Python fixture of handlers, services and a SQLite store, eleven rules and no real repositories: results need replication with stronger models, other languages and larger codebases.
Stronger models may violate rules less often without enforcement, which would leave less room for enforcement to help; this is the main limit on generalisation, and low-cost models were chosen because the comparison needs violations to occur, not because the conclusion is claimed for them alone.
The study was run by one independent researcher without dedicated funding, which shaped its scale but not its design.
\textbf{Conclusion.} Few informative tasks and few repetitions per cell; only one of the confirmatory tests was significant; Study~1 is exploratory, with nine contrasts none of which was significant after correction; effects of models and tasks are confounded across studies; the functional non-inferiority result is trivial where all runs pass.
Study~1 sent prompts to a provider's unpaid tier, whose terms state that submitted content may be used to improve the provider's products and read by human reviewers; the prompts contained only task text and synthetic code, and no personal data. The daily quota of that tier was handled by rotating among several API keys, which does not affect the outputs but may be read as conflicting with the provider's terms on documented usage limits; Studies~2 and~3 used a paid interface and no rotation.

\section{Conclusion}
Across three studies on two model families, enforcing engineering contracts during agent execution produced no oracle-detected violation in 104 paired runs and did not reduce functional correctness in the confirmatory studies, whereas the same rules as prompt text left 14 to 50\% of runs violating and checks after generation could not repair commands that had already run.
In the targeted confirmatory study the advantage over no rules was significant (31 points, Holm $p=0.014$); the advantage over prompted rules was in the same direction (17 points) but not significant, and Study~2 could not resolve either contrast.
These results show that enforcement before execution prevents violations that prompting and post-hoc checking do not, for the rules and tasks we tested; they do not show that prompting is useless, since it removed some violation types entirely, nor that the effect holds for stronger models or realistic repositories.
Two practical lessons follow from the failures we found: rule text and the permissions in the surrounding prompt must not conflict, and an enforced rule needs an independent check and human review, because the gate and its oracle can share a blind spot.
Future work should test stronger models and larger repositories, measure the false-block rate against human judgement, widen the rule set, and replicate the comparison of prompt text, post-generation checks and in-execution enforcement with a pre-registered protocol.

\section*{Data and Artifact Availability}
The tasks, hidden tests, reference solutions, oracles, rule library, harness configuration, run logs, scoring and analysis scripts, review packets and the frozen plans with their hashes are released at \url{https://github.com/RH2004/Beyond-Prompts-Runtime-Contracts-for-Coding-Agents}. Code is licensed under the PolyForm Noncommercial License 1.0.0 and data and documents under CC BY-NC 4.0; commercial use requires a separate written license from the author. No API keys are included.

\section*{Acknowledgments}
The author thanks two anonymous reviewers for the blind review of the 30 packets. The research received no external funding.

\end{document}